# Resonant tunneling through double-barrier structures on graphene[*]

Deng Wei-Yin(邓伟胤)[a)], Zhu Rui(朱 瑞)[a)], Xiao Yun-Chang(肖运昌)[b)] and Deng Wen-Ji(邓文基)[a)†]

*a) Department of Physics, South China University of Technology, Guangzhou 510640, China*

*b) LQIT, ICMP and SPTE, South China Normal University, Guangzhou 510006, China*

**Abstract:** Quantum resonant tunneling behaviors of the double-barrier structure on graphene are investigated under the tight-binding approximation. The Klein tunneling and resonant tunneling are demonstrated for the quasiparticles with energy close to the Dirac points. Klein tunneling vanishes by increasing the height of potential barriers to more than $300 \text{meV}$. The Dirac transport properties continuously change to the Schrödinger ones. It is found that the peaks of resonant tunneling approximate with the eigen-levels of graphene nanoribbons under appropriate boundary conditions. Comparison between the zigzag- and armchair-edge barriers is given.

**Key words:** Graphene, Tight-binding approximation, Resonant tunneling.
**PACS:** 72.80.Vp, 73.23.Ad, 73.40.Gk

1. Introduction

In recent years, graphene has been paid an extensive study of its physical properties both theoretically and experimentally [1-3]. The low-energy electronic states in graphene can be described by the Dirac equation while their high-energy counterpart obeys usual Schrödinger physics. Because of its linear dispersion relation at low energies, many unusual physical behaviors have been found, such as the Klein paradox [4], the conductance minimum [5], the unconventional quantum Hall effect [6-8], the shot noise with the Fano factor close to $1/3$ [9-11], sign reverse in quantum pumping [12, 13] and the specular Andreev reflection in the superconductor junction [14-16]. Based on these extraordinary properties, graphene has a lot of potential applications, such as transistors [17, 18], strain sensors [19], electrodes [20], entanglement source [21, 22] and etc.

---

[*] This project was supported by the National Natural Science Foundation of China (No 11004063) and the Fundamental Research Funds for the Central Universities, SCUT (No 2012ZZ0076).

[†] Corresponding author. E-mail: phwjdeng@scut.edu.cn

One of the proposed devices is the graphene-based single- and multi-potential-barrier structures [23-28]. Katsnelson et al. calculated the resonant conditions of transmission in the single barrier system, and found the resonant Fermi energy satisfies $E = V_0 \pm \left[ (N\pi/d)^2 + k_y^2 \right]^{1/2}$, where $N = 0, \pm 1, \cdots$, and $V_0$ is the height of the barrier [4]. Bai et al. studied the transport properties in graphene superlattices and found that the normal-incidence transmission resonances occur in the bilayer graphene structure but not in the monolayer one as a result of the Klein paradox [23]. The resonant tunneling through graphene-based double barriers has been studied by Pereira et al., they analyzed the origin of the resonance effect in this system and found the resonant features result from resonant electron states in the wells or hole states in the barriers [24]. However, most of the previous works focuses the low-energy Dirac regime and leaves the high-energy Schrödinger regime and that in between untouched.

In this Letter, we revisit the resonant tunneling processes in the graphene-based double-barrier structure. The tight-binding approach we used works in the whole energy regime beyond the low-energy Dirac-equation approximation. Previous published results can be recovered. Here, we mainly focus the regime where the potentials are so strong that in the barrier regions the electronic states can't be described by the Dirac equation. Continuous transition from the Klein tunneling is also discussed. So the considered is a hybrid system where in the non-barrier regions the electronic states are described by the Dirac equation while in the barrier regions they can be described by both the Schrödinger and Dirac equations dependent on the barrier heights.

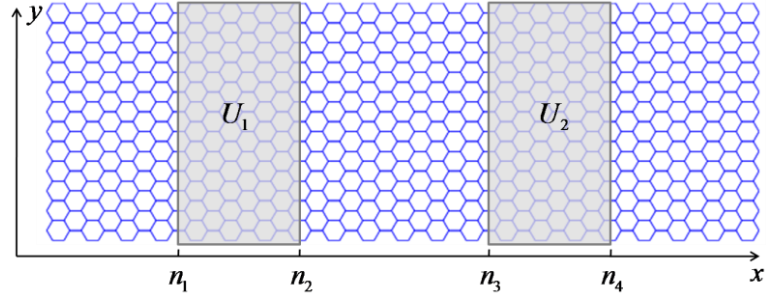

Fig. 1. Bird's view sketch of the double-barrier structure in graphene.

2 Model and Formulation

We consider the double-barrier model in an undoped graphene sheet occupying the $x$-$y$ plane, in which the left and right parts are two semi-infinite leads separated by the electric gates' modulated region. The geometry sketch is shown in Fig. 1. The grown direction is taken as the $x$ axis and the width along $y$ direction is infinite. The left lead extends from $x=-\infty$ to $x=n_1 l$, where $l=\sqrt{3}a_0/2$ and the lattice constant [29] $a_0=0.246$nm; the left barrier region contains $N_1=n_2-n_1$ zigzag chains so the width of this region is $N_1 l$ and the right barrier region contains $N_2=n_4-n_3$ zigzag chains; the middle region contains $M=n_3-n_2$ zigzag chains so the width is $Ml$; and the right lead occupies $x>n_4 l$.

According to the tight-binding approximation, the corresponding Schrödinger equation is

$$\varepsilon_n \varphi_n + \sum_{\delta} t_{n,n+\delta} \varphi_{n+\delta} = E\varphi_n, \qquad (1)$$

where subscript $n$ is the ordinal number of the atoms, $\varphi_n$ denotes the wave function in the Wannier representation, $\varepsilon_n$ and $t_{n,n+\delta}$ denote the site energy and the nearest-neighbor hopping integral, respectively. In perfect graphene the structure is composed of two sublattices, labeled by $A$ and $B$. In presence of a 1D confining potential $U=U(x)$, the lattice wave function can be written as $\varphi_{n,m-1}^{A,B}=e^{-ia_0 k_y}\varphi_{n,m}^{A,B}$

due to the translational invariance along the $y$ direction. Substituting it into Eq. (1), we have

$$-\psi_{n-1}^A + (\varepsilon_n - E)\psi_n^B - \tau\psi_n^A = 0,$$
$$-\tau\psi_n^B + (\varepsilon_n - E)\psi_n^A - \psi_{n+1}^B = 0.$$
(2)

where $\psi_n^B = \varphi_{n,m}^B$, $\psi_{n+1}^B = e^{ia_0 k_y/2}\varphi_{n+1,m}^B$, $\psi_n^A = e^{ia_0 k_y/2}\varphi_{n,m}^A$, $\psi_{n+1}^A = \varphi_{n+1,m}^A$ and $\tau = 2t_0 \cos(a_0 k_y/2)$. The gated site energies can be written as

$$\varepsilon_n^B = \begin{cases} \varepsilon_0 + U_1, & n_1 < n \leq n_2, \\ \varepsilon_0, & \text{others}, \\ \varepsilon_0 + U_2, & n_3 < n \leq n_4, \end{cases} \quad \varepsilon_n^A = \begin{cases} \varepsilon_0 + U_1, & n_1 \leq n < n_2, \\ \varepsilon_0, & \text{others}, \\ \varepsilon_0 + U_2. & n_3 \leq n < n_4. \end{cases}$$
(3)

We define the dimensionless variables $U_j \to U_j/t_0$ and $x \to x/l$ and set the site energy of the gate-free regions $\varepsilon_0 = 0$. The homogeneous hopping integrals are $t_{n,n+\delta} = -t_0 = -2.7\text{eV}$. Hence solution of incidence from the left reservoir can be written as

$$\psi_n^B = \begin{cases} e^{i\kappa(n-n_1)} + re^{-i\kappa(n-n_1)}, & n \leq n_1 \\ A_2 e^{i\kappa_1(n-n_1)} + B_2 e^{-i\kappa_1(n-n_1)}, & n_1 < n \leq n_2 \\ A_3 e^{i\kappa(n-n_2)} + B_3 e^{-i\kappa(n-n_2)}, & n_2 < n \leq n_3 \\ A_4 e^{i\kappa_2(n-n_3)} + B_4 e^{-i\kappa_2(n-n_3)}, & n_3 < n \leq n_4 \\ te^{i\kappa(n-n_4)}, & n > n_4 \end{cases}$$
(4)

$$\psi_n^A = \begin{cases} ae^{i\kappa(n-n_1)} + a^* re^{-i\kappa(n-n_1)}, & n < n_1 \\ a_{1+} A_2 e^{i\kappa_1(n-n_1)} + a_{1-} B_2 e^{-i\kappa_1(n-n_1)}, & n_1 \leq n < n_2 \\ aA_3 e^{i\kappa(n-n_2)} + a^* B_3 e^{-i\kappa(n-n_2)}, & n_2 \leq n < n_3 \\ a_{2+} A_4 e^{i\kappa_2(n-n_3)} + a_{2-} B_4 e^{-i\kappa_2(n-n_3)}, & n_3 \leq n < n_4 \\ ate^{i\kappa(n-n_4)}. & n \geq n_4 \end{cases}$$
(5)

where $\kappa, \kappa_1, \kappa_2$ are the dimensionless momentums along the $x$ direction in the normal, left-barrier and right-barrier regions, respectively. They are determined by the dispersion relations

$$E = \pm |\tau + e^{i\kappa}|,$$
$$E = U_1 \pm \sqrt{\tau^2 + \tau(e^{i\kappa_1} + e^{-i\kappa_1}) + 1}, \quad (6)$$
$$E = U_2 \pm \sqrt{\tau^2 + \tau(e^{i\kappa_2} + e^{-i\kappa_2}) + 1}.$$

And $a = -(\tau + e^{i\kappa})/E$, $a_{1\pm,2\pm} = -(\tau + e^{\pm i\kappa_{1,2}})/(\varepsilon - U_{1,2})$, $a^*$ is complex conjugate of $a$; $A_i, B_i (i = 2,3,4)$ are amplitudes of the wave functions traveling along the $x$ and $-x$ direction of the $i$-th region.. Then $r$ and $t$ are the reflection and transmission amplitudes, which can be determined by matching the eigen equations at region interfaces which can be transformed to wave function continuity at the interfaces.

Compared with the ab initio calculation we know that the nearest-neighbor tight-binding model in graphene is valid at energies below 2eV [29], which is much higher than the energy of Dirac-like particles ($<$1eV [4], actually the energy should be less than 300meV, which we will discuss below). We choose $\vec{K}$ point ($\vec{K} = 2\pi(0,2/3)/a_0$) as the origin and the incident angle $\theta$ spans between $k_x$ and $k_y$. When the Fermi energy agrees with the allowable energy of the structure, resonance effect occurs. The resonance effect in this structure contains the quasiparticle resonance in the barriers, the resonance in the middle region and the hybridization between them. When the Fermi energy is in the barrier, resonance comes from hole eigen-states, while that comes from the electron ones in the middle region. In our model, the barriers and the middle region are infinite-length zigzag nanoribbons. The edge state energy is very small compared to the Fermi energy when the nanoribbon is wide enough so it does not reshape the resonance effect. Dispersion of the standing wave state in the zigzag nanoribbon [30] is

$$\varepsilon_\pm = \pm |1 + \tau e^{i\kappa}|, \quad (7)$$

and the wave vector can be determined from the relation

$$\frac{1 + \tau e^{i\kappa}}{|1 + \tau e^{i\kappa}|} e^{iM\kappa} = \pm 1, \quad (8)$$

where $M$ is the width of the nanoribbon. When the Fermi energy is equal to one of the allowable energies in the zigzag nanoribbon, namely $E = \varepsilon_\pm$, resonance effect occurs. For normal accidence, $\tau = -1$ and $\kappa = \pi(1+2n)/(2M+1)$, $n = 1, 2, \cdots, M$. When the incident energy is small, or $M$ is very large, $\kappa$ is small. Eigen energies (7) can be approximated into

$$\varepsilon_\pm = \pm\sqrt{2(1-\cos\kappa)} \approx \pm\kappa \approx \pm\left(\frac{\pi}{2M} + \frac{\pi}{M}n\right). \tag{9}$$

Separation between two transmission resonances is $\pi/M$. It is difficult to obtain the analytical expression of allowable energies for large incident angles, because of the anisotropic momentum space. If the barriers and the middle region are Armchair nanoribbons along the $x$ direction, the resonance energy is a little different from the above system. In this case the vector in the confined direction $k_y = \left[\sqrt{3}n\pi/(M+1) + 2\pi/\sqrt{3}\right]$ which is not related to the vector $\kappa$ in the infinite direction [31, 32]. The first order approximate allowable energy of the Taylor expansion is zero and the second one can be obtained

$$\varepsilon_\pm \approx \pm\sqrt{\left(\frac{\sqrt{3}n\pi}{M+1}\right)^2 + \kappa^2}. \tag{10}$$

$(M+1)/\sqrt{3}$ is the dimensionless width of the Armchair nanoribbon. This is the same as resonance conditions of the low energy situation because of the similar structure of the momentum space where the vectors of the two directions are independent [4].

With the obtained transmission coefficients, the Landauer-Büttiker conductance can be calculated as [33, 34].

$$G(E_F) = G_0 \int_{-\pi/2}^{\pi/2} T(E_F, k_F \sin\theta)\cos\theta d\theta. \tag{11}$$

where $\theta$ is the incident angle relative to the $x$ direction, $G_0 = 2e^2 k_F L_y/(\pi h)$ is taken as the conductance unit, $L_y$ is the sample size along the $y$ direction and $k_F = \sqrt{\kappa^2 + k_y^2}$.

## 3 Numerical Results and Discussion

In the numerical calculations, we take $U_1 = U_2 = U$. The transmission $T(E, \theta, U, N_1, M, N_2)$ and conductance $G$ are functions of the Fermi energy, the incident angle, the strength of the double potentials and the widths of barriers and the middle region.

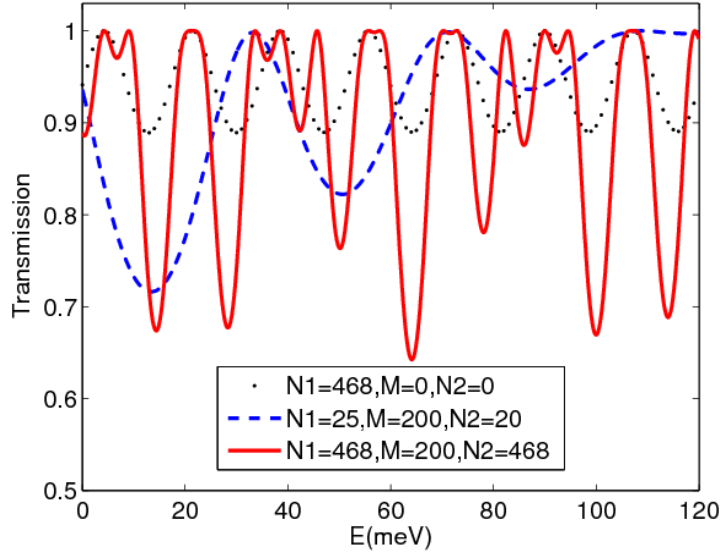

Fig. 2. Transmission probability of normal incidence as a function of the Fermi energy. The potential strength is $U = 1800\text{meV}$.

We calculated the transmission as a function of the Fermi energy for different $N_j (j=1,2)$ and $M$ at normal incidence ($\theta = 0$), shown in Fig. 2. The black dotted line, blue dashed line and red solid line correspond to $(N_1, M, N_2) = (468, 0, 0)$, $(25, 200, 20)$ and $(468, 200, 468)$, respectively. From Fig. 2, we find three noteworthy features. First, there is no Klein tunneling with perfect transmission at all energies in normal incidence. The potentials are so high that the Dirac equation is not satisfied in the barrier regions. Second, for $(N_1, M, N_2) = (468, 0, 0)$ or $(25, 200, 20)$, separation between neighbored transmission resonance peaks is equal, which can be seen in Eq. (9). That is different from the resonance effect in the free electron gas, which has $n^2$ relation to the incident energy. Third, the resonance effect in this system contains resonance in the barriers, resonance in the middle region and the hybridization between them. This property is nontrivial since resonance in the barriers is a result of hole quasibound levels, which does not occur in semiconductor superlattices with band gaps forming the potential barriers. For $(N_1, M, N_2) = (468, 0, 0)$, the resonance effect is in the single barrier; for $(N_1, M, N_2) = (25, 200, 20)$, the resonance effect is in the middle region. The width of the barriers is so small that the bound level interval is so large that the resonance effect in barriers is invisible in the figured energy range, and the unequal width can more suppress the resonance effect. For $(N_1, M, N_2) = (468, 200, 468)$, the resonance peaks contain contributions from quasibound states in the barriers as well as the middle region. The extra resonance peaks are the hybridization of the three regions.

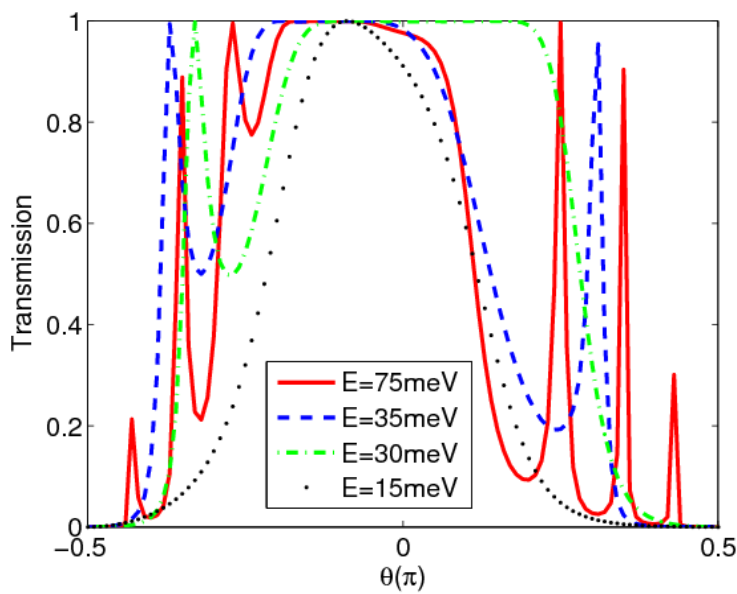

Fig. 3. Transmission probability as a function of the incident angle for different Fermi energies. Other parameters are $U = 1500 \text{meV}$, $N_1 = N_2 = 234$, $M = 468(100\text{nm})$.

Fig. 3 shows results of the transmission as a function of the incident angle for different Fermi energies $E = 75\text{meV}, 35\text{meV}, 30\text{meV}, 15\text{meV}$, corresponding to red solid, blue dashed, green dash-dotted and black dotted lines, respectively. All of the four lines demonstrate the same feature of incident-angle asymmetry. This can be intuitively understood from the dispersion relation of infinite graphene around the $\vec{K}$ point. The energy is asymmetric about $k_x$ for large quasiparticle energy when the dispersion relation is not linear. At this time the wave vector of angle $\theta$ is different from that of angle $-\theta$ at the same incident energy. This leads to the asymmetry of transmission about the incident angle. Among the four lines, the other shared feature is that perfect transmission occurs at some particular incident angles as a result of resonance effect [24]. At these incident angles, the Fermi energy is equal to the allowable energy of one of the zigzag nanoribbons. In addition, the resonant peak number increases with the Fermi energy.

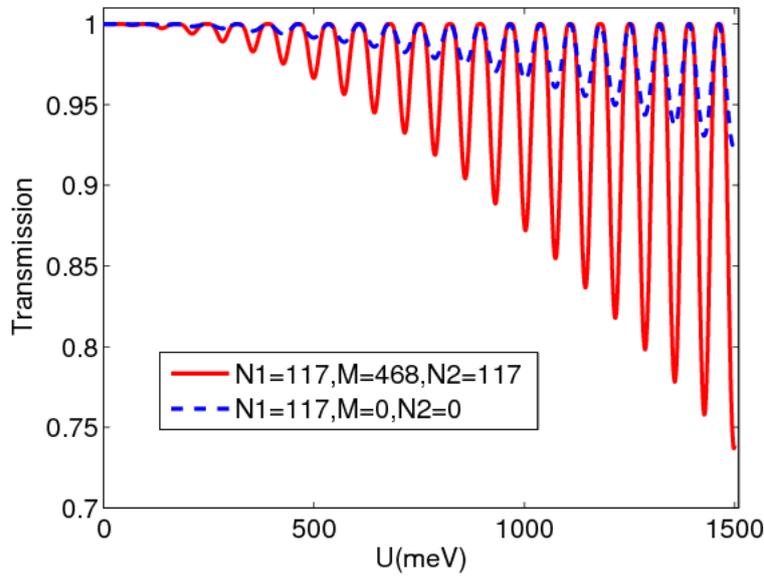

Fig. 4. Transmission probability as a function of the potential strengths. The Fermi energy is $E=100\text{meV}$.

In Fig. 4, normal-incidence transmission is plotted as function of the potential strength for a double-barrier (red solid line) and for a single-barrier (blue dashed line) structure. When the potential strength is small perfect transmission occurs with $T=1$ featuring the Klein paradox. With the potential height increased, the maximum transmission diminishes when the quasiparticle state in the barrier region should be described by the Schrödinger equation and the tunneling is not the Klein-type any more. From Fig. 4 it can be seen that the approximate upper threshhold of the Dirac equation in graphene is less than $300\text{meV}$. The periodic resonance peaks result from the resonance effect of the barriers. In the barrier region the resonance condition is

$E-U=\varepsilon_{\pm}$. So as the potential height increases, the transmission periodically oscillates.

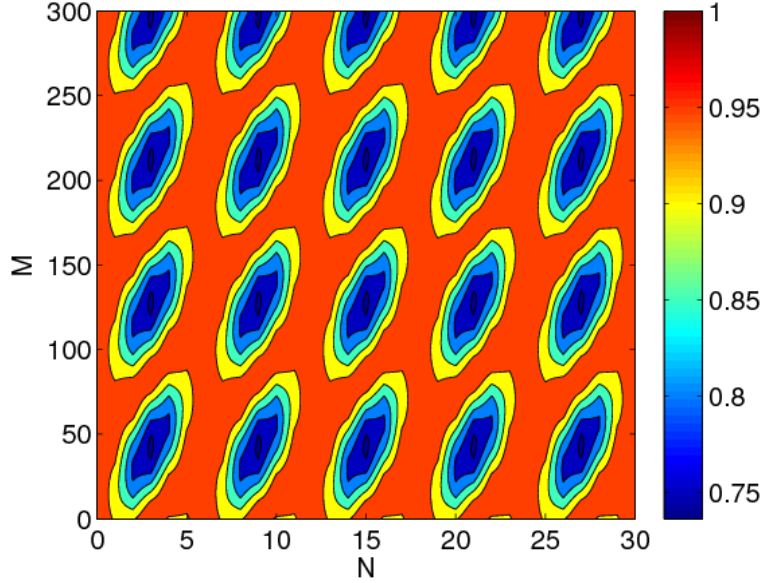

Fig. 5. Contour plot of the transmission probability. The parameters are $E=100\text{meV}$, $U=1500\text{meV}$, $N_1=N_2=N$.

In Fig. 5 we plot a $(N,M)$ contour of the transmission $T$ in the normal incidence. Roughness of the contour edges is due to discreteness in $N$ and $M$, which makes sense only with integer values. It naturally recovers smoothness in its equivalent continuous model. As seen, the transmission oscillates as a function of the width $N$ of the barriers region with a period. This is the resonance effect in the barriers. We can obtain resonant peaks $N=\left[\pi(2n+1)/\kappa-1\right]/2$, $n=1,2,\cdots,N$ from Eq. (8), where $\kappa\approx\pm(E-U)$. $T$ as a function of the width $M$ of the middle region is also periodic by the resonance effect in the middle region. The separation of resonant peaks of $N$ is shorter that $M$ because of potential $U$. In addition, when $N=0$, the transmission is perfect because there is no barriers in the system.

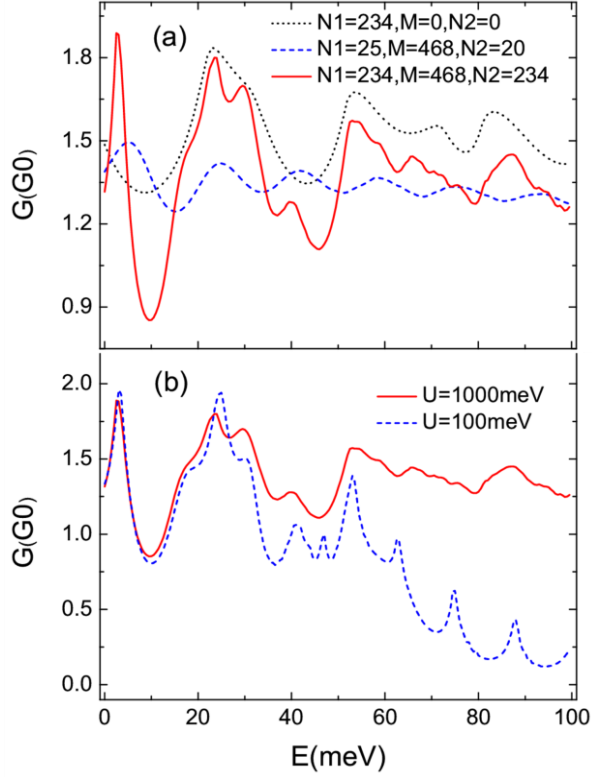

Fig. 6. Conductance as a function of the Fermi energy.

In Fig. 6 we plot the ballistic conductance as a function of the Fermi energy for different structure parameters. In Fig. 6 (a) the black dotted, blue dashed and red solid lines correspond to the barrier and middle regions with $(N_1, M, N_2) = (234, 0, 0)$, $(25, 468, 20)$ and $(234, 468, 234)$, respectively. The parameter used in the calculation is $U = 1000\text{meV}$. The resonance peaks of conductance result from the incident-angle averaged resonance effect of the barriers, middle region and the hybridization between them of these three regions. The resonant peaks are not equidistant because of the integral of all incident angles. In Fig. 6 (b) the red solid and blue dashed lines correspond to the potentials with $U = 1000\text{meV}$ and $100\text{meV}$, respectively. Other parameters are $N_1 = N_2 = 234$, $M = 468$. We can find that the conductance between them is similar. So resonance effect has a significant influence on the tunneling conductance even in the low energy. Actually, the Klein tunneling plays an important role in transmission of normal incidence. For other incident angles the transmission is a combined effect of the Klein paradox and the resonance effect. Another feature is that the high-barrier conductance at larger Fermi energies can be significantly larger than that of small barriers. In the former case, more resonant levels contribute to the conductance as the Fermi energy increases.

4 Summary

Based on the tight-binding approximation, we have investigated the resonant tunneling properties through double-barrier structures on graphene. The resonant transmission as functions of the Fermi energy, the incident angle, the potential strength, the widths of the barriers and the middle region were calculated, and the influence of the resonant tunneling to the conductance was discussed. It is shown that the resonance effect in this system contains the resonance in the barriers, the resonance in the middle region and the hybridization between them, and the resonant peaks are equidistant in the normal incidence. Compared to the low barriers, the conductance of high-barrier structures can be significantly increased by more resonant levels contribute.